\documentclass[aps,pra,twocolumn,showpacs,amsmath,amssymb,preprintnumbers,superscriptaddress,10pt]{revtex4-1}

\usepackage{graphicx}
\graphicspath{{figures/}}
\usepackage{SIunits}
\usepackage{hyphenat}
\usepackage[bookmarks=false]{hyperref}
\usepackage{color}
\usepackage[capitalise]{cleveref}
\crefname{section}{Sec.}{Secs.}% APS style uses abbreviations
\Crefname{section}{Section}{Sections}

\definecolor{pink}{RGB}{255,0,255}

\begin{document}

\title{Homodyne-detector-blinding attack in continuous-variable quantum key distribution} 

\author{Hao~Qin}
\email{qinhao@casquantumnet.com}
\affiliation{Institute for Quantum Computing, University of Waterloo, Waterloo, ON, N2L~3G1 Canada}
\affiliation{\mbox{Department of Combinatorics and Optimization, University of Waterloo, Waterloo, ON, N2L~3G1 Canada}}
\affiliation{Department of Physics and Astronomy, University of Waterloo, Waterloo, ON, N2L~3G1 Canada}
\affiliation{CAS Quantum Network Co.,\ Ltd.,\ 99 Xiupu road, Shanghai 201315, People's Republic of China}

\author{Rupesh~Kumar}
\affiliation{Department of Physics, University of York, York YO10~5DD, UK}

\author{Vadim~Makarov}
\affiliation{Russian Quantum Center and MISIS University, Moscow, Russia}
\affiliation{Department of Physics and Astronomy, University of Waterloo, Waterloo, ON, N2L~3G1 Canada}

\author{Romain~All\'eaume}
\affiliation{T\'el\'ecom ParisTech, LTCI, 46 rue Barrault, 75634 Paris Cedex 13, France}

\date{\today}

\begin{abstract}
We propose an efficient strategy to attack a continuous-variable quantum key distribution (CV-QKD) system, that we call homodyne detector blinding. This attack strategy takes advantage of a generic vulnerability of homodyne receivers: a bright light pulse sent on the signal port can lead to a saturation of the detector electronics. While detector saturation has already been proposed to attack CV-QKD, the attack we study in this paper has the additional advantage of not requiring an eavesdropper to be phase locked with the homodyne receiver.  We show that under certain conditions, an attacker can use a simple laser, incoherent with the homodyne receiver, to generate bright pulses and  bias  the excess noise to arbitrary small values, fully comprising CV-QKD security. These results highlight the feasibility and the impact of the detector blinding attack. We finally discuss how to design countermeasures in order to protect against this attack.
\end{abstract}

\maketitle

\section{Introduction} \label{Introduction}	
Quantum Key distribution (QKD) \cite{Scarani2009,Gisin2002} is one of the most important and practical applications of quantum information processing. It has already been made commercially available and has deployed in test and production environments. QKD allows two remote parties Alice and Bob, to establish a secret key over a public quantum channel, assisted by a classical communication channel. QKD security can notably be guaranteed even against computationally unbounded adversaries, making QKD the only available information-theoretic secure key establishment scheme practically available to date, beyond the use of physically-protected secret couriers. QKD information-theoretic security is however relies on some minimal set of assumptions: Alice and Bob labs, where secret information is stored and processed, should not leak this information to the outside world, moreover, Alice and Bob hardware (laser, modulators, detectors) are supposed to behave (at least approximately) according to an abstracted model, that then allows to proof theoretical security. However, in practice, the real-world QKD implementations not act exactly verify the aforementioned assumptions and such deviations may lead to vulnerabilities and enable an eavesdropper, Eve, to launch so-called side channel attacks and break the security of practical QKD devices.
 
In discrete-variable (DV) QKD, single photon detector (SPD) is the most exposed device, from the implementation security viewpoint, and several attack strategies have been proposed to exploit SPD vulnerabilities and attack DV-QKD implementations. Attacks such as time shift \cite{Qi2007,Zhao2008}, after gate \cite{Wiechers2011}, blinding \cite{Makarov2009}, spatial mode mismatch \cite{Sajeed2015a} attacks and etc. may all lead to security breach. Among those attacks, the blinding attack is probably considered as the most powerful attack, as this attack strategy allows Eve to actively control Bob's SPD remotely, using intense light. Such kind of attack has been experimentally demonstrated on commercial QKD systems \cite{Lydersen2010} and in a full-field deployment \cite{Gerhardt2011}. Various countermeasures \cite{Huang2016a,Lydersen2010a} have been proposed against detector-based attacks. However only measurement-device-independent (MDI) QKD \cite{Lo2012,Braunstein2012}, i.e. QKD protocols where security can be established without trusting the detector, can firmly demonstrate to be immune against these attacks targeting SPDs.

Continuous-variable QKD (CV-QKD) \cite{Weedbrook2012a}, is another promising approach to perform quantum key distribution. It relies on continuous modulation of the light field quadratures and measurements with coherent detection (homodyne or heterodyne detectors) instead of SPDs in DV-QKD system. Benefiting from coherent detection, CV-QKD can be fully implemented with off-the-shelf optical communication components \cite{Lodewyck2007,Fossier2009,Jouguet2013a}. Moreover, the local oscillator (LO) in the coherent detection acts as a "built-in" filter to efficiently remove any noise photons in different modes, which enable CV-QKD to be deployed in co-existence with intense classical channels over optical networks \cite{Kumar2015} and to be possibly implemented in day light free space environments. CV-QKD practical implementations however also suffer from potential vulnerabilities. For example LO manipulation is a long standing security problem: if LO is sent on the public channel, then an attacker can modify LO pulses in different ways \cite{Jouguet2013,Ma2014, Ma2013a,Huang2013,Huang2014} and learn secret keys without being discovered. A generic solution to this issue has however been recently proposed: by generating locally  the LO (LLO) pulses at Bob side \cite{Qi2015, Soh2015, Huang2015a, Marie2017}. Regarding the homodyne detection (HD), which is a central component in CV-QKD, it has been shown in  \cite{Qin2013,Qin2016}, under the name ``saturation attack''  that HD saturation can be induced by an attacker and exploited to launch attacks that can fully break CV-QKD security. More precisely it has been shown that HD saturation induced by a coherent displacement can bias the excess noise estimation and conceal the presence of an eavesdropper, performing intercept-resend attack on the signals sent by Alice.  However, coherently displacing the signal sent by Alice, without adding detectable excess noise is highly challenging, making this attack strategy difficult to implement in practice. 

In this paper, inspired by the blinding attack in DV-QKD, we propose a simple and practical way to saturate a homodyne detector with finite linear detection range. The attack exploits the loss imbalance of the two ports, in a balanced HD and consists in sending a bright pulse onto the signal port, to induce electronic saturation. Such loss imbalance is quite generic to any HD implementation: the two photodiodes quantum efficiencies as well as the beam-splitter reflection/transmittance coefficients are never perfectly balanced. This implies the need for additional balancing, that is in general ensured by introducing some variable attenuation in one of the optical arms. Such balancing must be done with precision with respect to the LO port, since LO pulses are intense. But as a consequence, a good balancing in practice cannot, in return, be guaranteed with respect to the other port, i.e. the signal port. As a consequence, any relatively strong light impinging on the signal port will produce a comparatively stronger photocurrent on one of the HD photodetectors, which will further cause HD's amplifier electronic saturation.

Due to HD saturation, Bob's HD output signals are limited within some finite range. This however violates a basic assumption generally used in  CV-QKD security proof: linearity, namely that Bob's HD output signal is supposed to be linearly proportional to the input optical quadratures. The principle of the blinding attack we introduce here consists, for Eve, to actively drive Bob's HD into a saturated response mode, by sending strong external pulses on the signal port. We will show that such manipulation can be used to reduce  the estimated excess noise, under certain conditions. More precisely, we will show  that combining the sending of strong light pulses to Bob, with a full intercept-resend attack, Eve can break the security of the widely used Gaussian Modulated Coherent State (GMCS) CV-QKD protocol \cite{Grosshans2002,Grosshans2003}. We will analyze the conditions such that Eve can achieve a full security break, illustrating that this attack can be implemented with current technologies and a low-complexity experimental system. Importantly, our attack only targets on the HD which means even the recent proposed LLO CV-QKD scheme is not immune to this attack if no countermeasure is considered. This highlights that detector loopholes also exist in CV-QKD and can potentially affect all CV-QKD implementations. Finally we will discuss possible countermeasures against detector-based attacks in CV-QKD and compare them with countermeasures against blinding attack in DV-QKD.

This paper is organized as follows. In \cref{secii}, we present the security basis of GMCS CV-QKD: parameter estimation and its relation to quantum hacking. In \cref{seciii}, we study experimentally several imperfections of a practical HD and predict the shot noise measurements with the proposed HD model. In \cref{seciv}, we introduce the attack strategy of the HD blinding attack. In \cref{secv}, we perform the security analysis of the proposed strategy and demonstrate its security breach feasibility in simulations. At last, we discuss possible countermeasures against HD blinding attack in CV-QKD in \cref{secvi}, and conclude in \cref{secvii}.

\section{Practical security in CV-QKD}\label{secii}
In this section, we briefly present the Gaussian Modulated Coherent State (GMCS) CV-QKD protocol. This protocol is widely used, notably thanks to a well-understood security proof, base on the optimality of Gaussian attacks \cite{diamanti2015distributing}. We will be focused on the GMCS protocol throughout the paper, and illustrate in this section the connection between the parameter estimation phase, and  the practical attack strategy. 

\subsection{GMCS protocol and parameter estimation}\label{GMCS}
In GMCS protocol~\cite{Grosshans2002}, Alice prepares the coherent state $|X+iP\rangle$ as the quantum signal, in which amplitude $X$ and phase $P$ quadratures are continuously modulated with a centered Gaussian distribution with a variance $V_AN_{0}$. The shot noise $N_{0}$ is the HD variance when the input signal is vacuum field. At Bob's side, he performs HD on Alice's signal by interfering it with strong phase reference LO. Bob randomly chooses to apply a phase modulation $0$ or $\pi/2$ on LO in order to measure quadrature $X$ or $P$ in phase space. Note that, it is not necessary for Alice to send LO over the insecure channel, Bob can generate the LO at his side and recover the phase information with help of additional reference pulses from Alice \cite{Qi2015,Soh2015,Huang2015a,Marie2017}. By repeating such process and sifting, Alice and Bob then obtain correlated Gaussian variables $ (X_A, X_B)$ as the raw keys. With reverse reconciliation~\cite{Grosshans2002,Jouguet2011}, Alice and Bob can extract an identical bits string from the correlated variables and obtain a secret key through privacy amplification. 

In order to estimate Eve's knowledges on the raw key and eliminate them in privacy amplification, an important step for Alice and Bob is to perform the parameter estimation to estimate excess noise, channel transmission and secret key rate. Security proofs of CV-QKD show that Gaussian attack is the optimal one which has been proven in collective attacks with asymptotic limit \cite{Garcia-Patron2006,Navascues2006}, in recent composable security proof\cite{Leverrier2015} and in arbitrary attacks with finite size \cite{Leverrier2017}. Such security proofs enable Alice and Bob to describe their quantum channel as a Gaussian linear channel which connects the raw data $X_A$ and $ X_B$ with a Gaussian noise factor $X_N$. This channel model allows Alice and Bob to determine the two characteristics of the quantum channel: excess noise $\xi$ and channel transmission $T$ by performing four measurements. Particularly, Alice's modulation variance $V_A$, Bob's HD variance $V_B$, Alice-Bob covariance $\text{Cov}_{AB}$ and shot noise calibration of Bob's HD variance $V_{B,0}$ when there is only LO input:
\begin{align}
\label{eq1}
V_{A}&=\langle X^2_{A} \rangle-\langle X_{A} \rangle^2,\\
\label{eq2}
\mathrm{Cov}_{AB}& = \langle X_AX_B\rangle-\langle X_A\rangle\langle X_B\rangle=\sqrt{\eta T}V_A,\\
\label{eq3}
V_{B}&=\langle X^2_{B} \rangle-\langle X_{B} \rangle^2 
=\eta T V_A + N_0 + \eta T \xi + v_\mathrm{ele},\\
\label{eq4}
V_{B_{0}}& = N_0 + v_\mathrm{ele},
\end{align}
in which $\eta$ and $v_\mathrm{ele}$ are Bob's HD overall efficiency and electronic noise which are calibrated before QKD, $N_0$ is the shot noise variance. Alice and Bob can  extract a portion of the raw key and estimate the channel transmission $T$ based on \cref{eq1,eq2}; and excess noise in shot noise units $\xi/N_0$ based on \cref{eq3,eq4}. They can then estimate the security key rate with a given security proof and decide whether to proceed to the key generation step or abort the protocol if there the secure key rate estimation is non-positive. Note that we need to take statistical fluctuation into account for the variance measurements with a realistic data block size $N$ \cite{Jouguet2012} in practice. In this paper, we want to emphasize the idea of the attack strategy, we only consider the collective attacks in asymptotic limit ($N\rightarrow \infty$).

\subsection{Quantum hacking in CV-QKD}
The goal of Eve's quantum hacking on CV-QKD system is to steal Alice and Bob's secret keys without being discovered. To achieve this, Eve is allowed to use every possible measure that is allowed by quantum mechanics to attack the open quantum channel. Some CV-QKD quantum hacking strategies such as wavelength attack \cite{Huang2013} and LO intensity fluctuation attack \cite{Ma2014} are only possible in theory that Eve has full access to future quantum computer with enough quantum memory. Under such cases, loopholes lead to increase of Eve's mutual information with Alice or Bob and to decrease the final secret key rate. It is however more important to study possible quantum hacking strategies in a realistic scenario when Eve's power is limited by current technologies, as it would bring immediate threats to CV-QKD security.

In CV-QKD,  excess noise estimation is the reference for Alice and Bob to decide to abort the protocol or proceed to key generation. Any flaw in the excess noise estimation can lead to security problem that Eve's attack action is undiscovered, which may fully compromise CV-QKD security. In order to attack CV-QKD with current technologies, Eve can perform an intercept-resend (IR) attack by using  optical heterodyne detection \cite{Lodewyck2007a} which corresponds to a ``entanglement breaking" channel. Under IR attack, Eve always has an information advantage over Alice and Bob but she also introduces at least $2N_0$ into their excess noise estimation  due to the heterodyne measurement disturbance and coherent state shot noise. Meanwhile, Eve can take advantage of CV-QKD implementation imperfections to formalize  particular attack strategies  to bias Alice and Bob's excess noise estimation in order to  hide her IR action and achieve a full security breach. For examples, in calibration attack \cite{Jouguet2013}, Eve delays the LO pulse such that Alice and Bob overestimate the shot noise based on their pre-established calibration, which results in underestimation of the excess noise. In saturation attack \cite{Qin2013,Qin2016}, Eve induces saturation on Bob's HD measurement and directly bias Alice and Bob's excess noise. In this paper, we also follow this idea and will take advantage of several imperfections in a HD to archive a  security breach on GMCS CV-QKD implementations.

\section{Imperfect homodyne detection in CV-QKD} \label{seciii}
In this section, we analyze HD imperfections such as imbalance and electronics saturation. These HD imperfections are the key elements that will be used in the HD blinding attack strategy.

\subsection{Practical homodyne detection with imperfections}\label{HDtheo}
\begin{figure}
	{\includegraphics[width=0.47\textwidth]{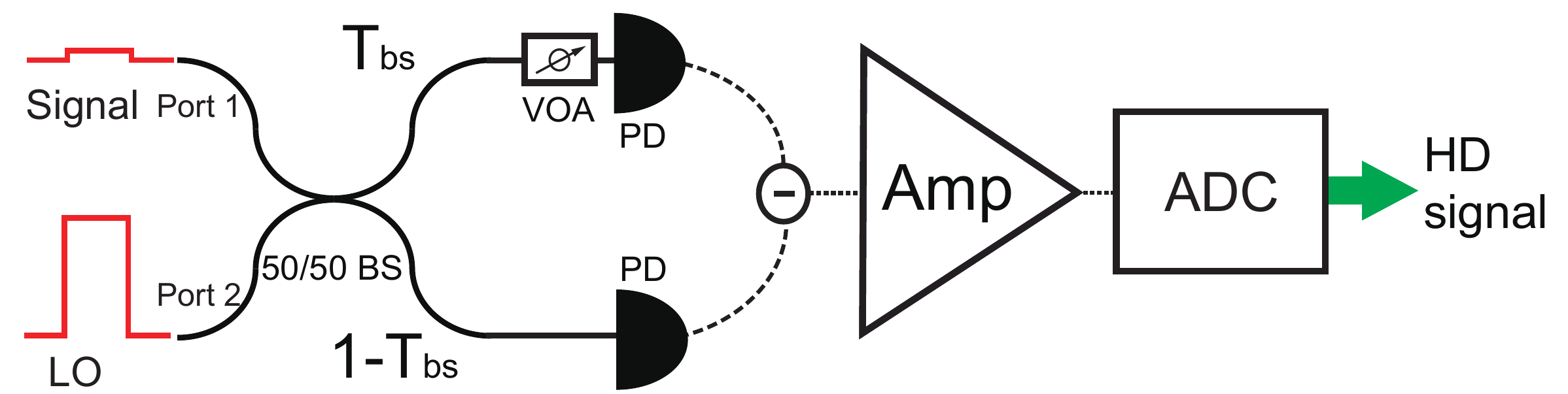}}
	\caption{ A simplified scheme of a practical HD, BS: beam-splitter, VOA: variable optical attenuator, PD: photodiode, AMP: amplifiers, ADC: analog-to-digital converter, solid line: optical signal, dash line: electronic signal.}
	\label{hd}
\end{figure}

In the context of CV-QKD , HD performance is usually measured by its overall efficiency $\eta$ and its electronic noise $v_\mathrm{ele}$ \cite{Lodewyck2007,Jouguet2013a}. However, imperfections such as limited bandwidth, linearity range, imperfect balance, etc., can also affect HD behavior and potentially impact on CV-QKD performance \cite{Kumar2012}. Some of these HD imperfections may even open security loopholes in CV-QKD to Eve which need to be carefully studied. Here, we are particularly interested in the imperfection of the HD on the optical part: imbalanced 50/50 beam-splitter (BS) and on the electronic part: finite linear range of detection.

% We first introduce a practical setup of HD, then we will analyze the impact of these two imperfections. 
As shown in \cref{hd}, a practical HD consists both optical and electronic parts. The input optical signals and LO pulses go into the two ports of 50/50 BS and with one another. Two output optical pulses then travel to two identical p-i-n photodiodes (PDs) which convert optical lights into two photocurrents with finite quantum efficiencies.  An electronic subtraction is then performed on the two  photocurrents and the subtracted photocurrent is amplified into a small voltage through a trans-impedance amplifier or a charge amplifier. The voltage signal is further amplified by a second stage amplifier to be detected by the Analog-to-digital converter (ADC) device (i.e., oscilloscope or data acquisition card). The final digitized data is  so-called HD output signal which is proportional to the input optical quadratures. The selection of the quadrature is dependent on the relative phase between LO and signal pulse. 

Thanks to the subtraction of HD, most LO intensities are eliminated while the rest energy carries the small quantum signal fluctuation which is ``amplified" by LO's amplitude. However due to the imbalance imperfection of the HD, a non-negligible leakage of LO contributes to the final HD output signal as an offset. Such leakage may also  contribute LO intensity fluctuation noise into HD  measurements if LO intensity is relatively high \cite{Chi2011}. In order to adjust the balance of HD, a variable optical attenuator (VOA) and a variable delay line (VDL) need to be added to one of the optical paths after the 50/50 BS. The balancing of HD is evaluated by the common-mode rejection ratio (CMRR) which is defined as $CMRR=-20 \log_{10}(2\epsilon)$ with $\epsilon$ as the overall imbalance factor  \cite{Chi2011}, $\epsilon$ quantifies the small deviation that varies from a perfect balanced HD. For example, a typical CMRR value of a well balanced HD is around $-52.4~\deci \bel$ \cite{Kumar2012} which means the difference between the two photocurrents before subtraction is $\epsilon=0.12\%$ over their total currents. 

% Note in CV-QKD, Bob's balancing setting is only valid for the LO pulse going into the LO port, but not for other light going into the signal port. 

In order to quantify the impact of such imbalance imperfection on HD, we analyze the case of shot noise measurement when there is no signal but only LO pulses sent into HD, in which we look at the first and second moment of HD statistics: mean and variance. To simplify the analysis, we consider the model of unbalanced HD with two ports in Ref.~\onlinecite{Ma2013a}. If there is only LO impinging on HD, the HD output state $X_{HD}$ can be given:
\begin{align}
\label{xhd}
X_{HD}=\eta(1-2T_\text{hd})I_\text{lo}+2\sqrt{\eta T_\text{hd}(1-T_\text{hd})I_\text{lo}}X_0+X_\text{ele},
\end{align}
in which $T_\text{hd}$ is the  overall transmission  of HD, which includes the  transmission of the 50/50 BS, optical loss in the optical path and efficiency of the PD while $1-T_\text{hd}$  is the overall   reflection, $\epsilon=1-2T_\text{hd}$ is thus the overall imbalance factor. $I_\text{lo}$ is the number of photons per one LO pulse (which is linear dependent on LO power or intensity), $X_0$ is the vacuum state, $X_\text{ele}$ is the HD electronic noise with a variance of $v_\mathrm{ele}$. We observe that the HD output is displaced by a value $D_\text{lo}$ [first term in \cref{xhd}] that is linearly proportional to $I_\text{lo}$  due to LO leakage, which directly determine the  mean of $X_{HD}$:
\begin{align}
\label{mhd}
\langle X_{HD} \rangle=D_\text{lo}=\eta (1-2T_\text{hd})I_\mathrm{lo},
\end{align}
here the vacuum state is centered on zero $\langle X_{0} \rangle=0$, and we assume the offset due to the HD electronics is small enough to be neglected $\langle X_\text{ele}\rangle\approx 0$. We can further deduce the HD variance based on \cref{xhd} with  the definition of variance:
\begin{equation}
\label{vhd}
\begin{split}
V_{HD}=&\langle X^2_{HD} \rangle-\langle X_{HD} \rangle^2 \\
=&\eta^2(1-2T_\text{hd})^2f_\text{lo}^2I_\text{lo}^2+4(1-T_\text{hd})T_\text{hd}\eta I_\text{lo}+v_\mathrm{ele},
\end{split}
\end{equation}
in which $f_\text{lo}=\sqrt{\langle I_\text{lo}^2\rangle-\langle I_\text{lo}\rangle^2}/I_\text{lo}$ is the intensity fluctuation ratio of LO over the measurement time  and  the first quadratic term of $I_\text{lo}$ is the noise variance due to LO intensity fluctuations \cite{Chi2011}. If we consider a  typical CV-QKD implementation \cite{Jouguet2013a} with a low LO power $I_\text{lo}$  as the order of $10^8$  and   HD is adjusted to be balanced ($T_\text{hd} \approx 0.5$), we can   neglect  the  LO intensity fluctuation noise and the degradation effect due to imbalance as  the factor $4(1-T_\text{hd})T_\text{hd} \approx 1$.  \cref{vhd}  can be  further simplified into:
\begin{align}
\label{vhd2}
V_{HD}\approx\eta I_\text{lo}+v_\mathrm{ele},
\end{align}
where the first term is known as the shot noise $N_0=\eta I_\text{lo}$ in CV-QKD, which is proportional to $I_\text{lo}$ and it can be interpreted as the HD signal variation due to interference between LO and the vacuum state with $\langle X^2_{0} \rangle =1$. The amplification of LO also applies to vacuum state which attributes to the term of $I_\text{lo}$ in \cref{vhd2}. As shown in \cref{xhd} and  \cref{vhd}, LO leakage due to HD imbalance contributes an offset of  $D_\text{lo}$ in HD output signals and associated LO intensity noises in the HD variance measurement.

Beside the HD imbalance imperfection, finite linear detection range of the electronics part can also influence HD measurements and may lead to security loopholes \cite{Qin2013,Qin2016}. An important assumption in CV-QKD is that Bob's HD measurement varies linearly with the input optical quadrature. However such assumption does not always hold in a practical HD, because if input field quadrature overpasses certain threshold, the corresponding photocurrent would be relatively large, which can saturate the electronics and results in saturation of HD output signal. Electronics saturations usually happen on the amplifier or on the data acquisition card (DAQ). Depends on the specific electronics design, the amplifiers usually saturate at few volts which  is the intrinsic characteristics of the electronics. DAQ detection range in CV-QKD is usually set to a small range (typically between $-1\volt$ and $1\volt$) to ensure its measurement step precision, in principle, this range can be set as large as possible but not infinite. However the overall linear detection range is still limited by the amplifier. Two p-i-n PDs in HD can also become saturated mainly due to screening of the electric field caused by photo-generated carriers \cite{Williams1992}. However such limit is often relatively high (e.g.,\ few mW for Thorlabs FGA01FC) and total optical power of LO and signal in CV-QKD system is much lower than this limit. Thus PD saturation is usually not the reason causes HD saturation and we consider this realistic assumption in this paper. In practice, HD saturation is unavoidable and it is important to make sure HD works in the linear region. HD saturation effect can be modeled by a simple HD model \cite{Qin2013,Qin2016} with upper and lower bounds $\alpha_1$ and $\alpha_2$, where Bob's HD output signal after ADC can be given as:
\begin{equation}
\label{xbs}
X_{HD_\text{r}}=\begin{cases}
\alpha_1, X_{HD} \geqslant \alpha_1\\
X_{HD}, \alpha_2<X_{HD}<\alpha_1\\
\alpha_2, X_{HD} \leqslant \alpha_2
\end{cases},
\end{equation}
in which $X_{HD}$ is given by \cref{xhd}. This model shows that the linearity range of HD is limited by [$\alpha_2, \alpha_1$], otherwise HD output signals will be saturated to the limits. The limits $\alpha_1$ and $\alpha_2$ need to be calibrated in practice and they are dependent on HD electronics as mentioned. Due to HD saturation, variations of HD signals become much lower compared to the case in linear detection region, which will affect the correctness of HD statistic measurements \cite{Qin2013,Qin2016}. Moreover, when there is also the offset due to imbalance imperfection on HD, it can significantly change the pre-calibrated linear detection ranges if the offset factor $D_\text{lo}$ becomes comparable to $\alpha_1$ or $\alpha_2$, which needs to be further studied in experiments.  

\subsection{Experimental analysis on a practical homodyne detector }
In order to study influences of  HD imbalance and electronics saturation on HD output signals, we design a simple experimental test on HD shot noise measurement. We slightly modify the standard shot noise measurement procedure and compare the results under different balancing settings. The key idea of this test is that we intentionally unbalance HD to study its influence on HD saturation limit and further on   HD output signals.

The experimental setup can be refereed to \cref{hd} where we only send LO pulses and then measure  HD signals. We use a $1550~\nano\meter$ distributed feedback (DFB) laser (Alcatel LMI1905) to first prepare a train of optical pulses with pulse widths 100 ns and repetition rate at 1 MHz as LO pulses. Our HD consists of two PDs (JDS Uniphase EPM 605), the AmpTek A250 as the  first stage amplifier with a charge amplifier setting and a MAX4107  as the second stage  amplifier. This HD features a low noise (with a noise variance at the order of $10 \milli \volt^2$) and a low bandwidth (about 10 MHz).   We send LO pulses into port 2 of HD (\cref{hd}) and roughly minimize the HD output by adjusting optical loss of one path in order to balance the HD, which is considered as the 1st HD balance setting.  After measuring the average optical power of the input pulses with a power meter, we then record HD signals over 1 second which corresponds to $10^6$ pulses. We adjust our DAQ (Model NI6111) detection  range to $[-0.5 \volt, 0.5 \volt]$ as the detection limits. Any HD signals out of this range will be saturated to $\alpha_1=0.5 \volt$ or $\alpha_2=-0.5 \volt$. Note this is not the saturation limit of our HD amplifier which is about $\pm 3 \volt$, but we want to limit the linear detection range to be small in order to highlight its influence on HD signals. Based on the measured $10^6$ HD signals, we then estimate the mean and variance of this set of data, which is considered as one shot noise measurement for a given LO intensity. With the same balance setting, we repeat this shot noise measurement by gradually increasing LO intensity. 
\begin{figure}
	{\includegraphics[width=0.5\textwidth]{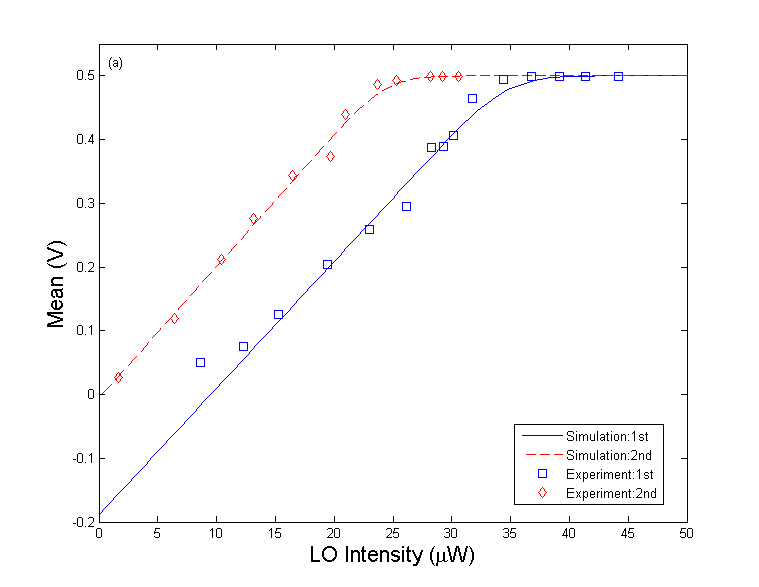}}
	{\includegraphics[width=0.5\textwidth]{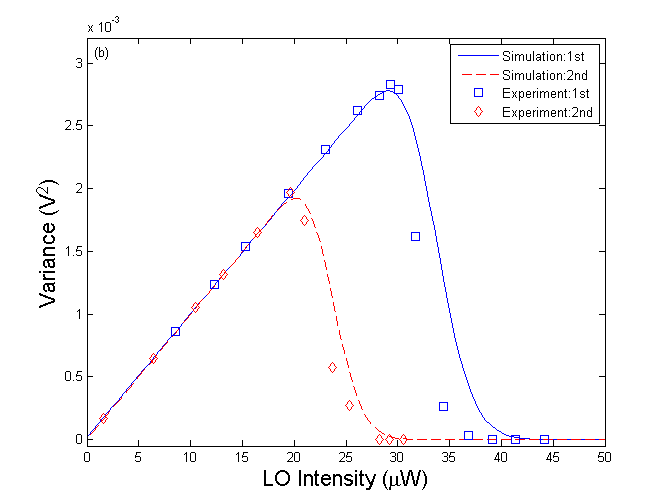}}
	\caption{Characterization of HD output voltage statistics, in absence of signal: under two different balancing conditions (solid lines and dashed lines). (a) Mean value ($V$) versus LO power. (b) Variance ($V^2$) versus LO power. }
	\label{mean2}
\end{figure}
The experimental results for the 1st setting are shown as blue squares in \cref{mean2}. As shown in \cref{mean2}(a), the mean value first increases linearly with respect to LO intensity, which matches the  prediction of \cref{mhd}. However, when the LO intensity reaches to about 35 $\micro \watt$, the mean value stops increasing but saturates at $0.5 \volt$. This is obviously due to the HD saturation effect predicted by \cref{xbs} when the imbalance offset $D_\text{lo}$ reaches to the detection upper limit $\alpha_1$. By applying \cref{vhd} into \cref{xbs} and fitting experimental data in the linear region, we successfully predict the behavior of HD mean values as shown on the blue solid curve in \cref{mean2}(a). 

Regarding to variance measurements, from \cref{mean2}(b) we can verify that HD variance increases linearly with LO intensity which matches the prediction of \cref{vhd}. Similarly, when LO intensity reaches to a relatively high value around 35 $\micro \watt$, the variance drops quickly and becomes  zero at about 45 $\micro \watt$. It is because the HD signals variation becomes much lower when HD is  saturated as any HD signal fluctuations beyond $\alpha_1=0.5 \volt$ have been cut off. We also observe that HD variance does not immediately turn into zero, because only parts of HD signals due to vacuum fluctuations have been limited by $\alpha_1$ between 35 $\micro \watt$ and 45 $\micro \watt$ LO intensity. We use the saturation model [\cref{xbs}] and shot noise variance [\cref{vhd}] to simulate HD variance behaviors as for the blue solid curve. As shown in \cref{mean2}(b), the simulation curve matches the behaviors of experimental HD variances data, which means we can account for the saturation model for further analysis.

In order to illustrate the impact of HD imbalance on HD signals, we now slightly adjust the optical loss of one path after BS to unbalance the HD. Such balance setting (2nd) imposes more LO leakage, which will further affect the behaviors of HD means and variances. With this balance setting, we repeat HD statistic measurements mentioned above and compare them with previous results at same LO intensity levels. Experimental and simulation results are shown as red diamonds and dashed curves in \cref{mean2}, respectively. As shown in \cref{mean2}(a), HD mean under 2nd setting reaches to saturation limit around $35 \micro \watt$ compared to the one at $45 \micro \watt$ in the 1st setting. It confirms \cref{vhd} that HD offset due to LO intensity leakage is proportional to $\epsilon$ and thus the equivalent displacements $D_\text{lo}$ on HD signals of the 2nd setting is larger than the one of the 1st setting. In consequence, HD signal of 2nd setting reaches to the detection limit $\alpha_1$ at a smaller value of LO intensity compared to the 1st setting, which can be observed by experimental (red diamonds) and simulation (red dashed curve) data in \cref{mean2}(b). The simulation curves also confirm the HD statistical behaviors, which shows that HD imbalance imperfections can influence the relation between HD saturation level and LO input intensity at a certain extent. Note in CV-QKD, HD is designed to precisely detect weak quantum signals, it can be saturated easily if the HD is not well balanced as LO intensity is usually many orders of magnitude stronger than quantum signal \cite{Chi2009,Fossier2009a}. 

These experimental and simulation results inspire us to  formalize a new attack strategy in CV-QKD similar to blinding attack in DV-QKD, where Eve inserts external lights into signal port of Bob's HD to influence HD output signals by taking advantage of HD imbalance and saturation imperfections. 
 
\section{Homodyne detector blinding attack on GMCS CV-QKD} \label{seciv}

\subsection{Principle of the attack }
Based on the previous analysis, Eve can formalize a simple strategy to saturate Bob's HD output signal by sending another incoherent classical light into HD's signal port instead of preparing coherent displacement as in saturation attack \cite{Qin2013,Qin2016}. Since Bob balances his HD with respect to the LO light that goes into LO port, any relatively strong light going into signal port cannot be subtracted as much as the one on LO port. Thus the external light contributes a strong offset on the HD output signal, at certain point it can cause HD saturation as shown in the previous section. In order to prevent inference with LO pulses, Eve can send the external lights in a different mode of the LO pulse, in practice she can use a different wavelength other than the one used for LO pulses. Moreover, due to the BS wavelength dependent properties, Eve has the possibility to ``control" the transmission of Bob's BS by selecting proper wavelength of the external light \cite{Huang2013,Ma2013a,Li2011}. On the other hand, the two PDs used in HD are classical detectors and usually have large wavelength ranges (typically from 800-1700 nm). Any lights in the sensitive range of the PD can produce photocurrents and contribute to final HD signals, which is impossible for Alice and Bob to distinguish the source of light by only measuring HD signals. However, as Eve's external light is incoherent with Alice and Bob's LO in CV-QKD, the external lights contribute excess noises into Alice and Bob's HD measurements. On the other hand, such excess noises due to the external lights can be ``sufficiently filtered" by the LO pulses,  in the sense that  external light is not interfered with LO and the related excess noise will be further normalized by a factor of $N_0$. 
 
As mentioned, in order to break the security with current technologies, Eve can combine this strategy with the IR attack. If Bob's HD works in the linear region, Alice and Bob can always notice the excess noise due to Eve's IR attack and the external light. However, Eve can always cause Bob's HD signal saturation by sending the external light strong enough. In this sense, Eve is able to ``control" Bob's HD signals and manipulate the HD statistic measurements. If Eve carefully selects the properties of the external light, her manipulation of Bob's HD signals can further lead Alice and Bob to underestimate the excess noises from Eve's actions which fully compromise the security. We will see that such attack strategy is simple to realize in experiments but powerful enough for Eve to steal keys without being discovered in the following sections. 
\begin{figure*}[t]
	\centering 
	{\includegraphics[width=1\textwidth]{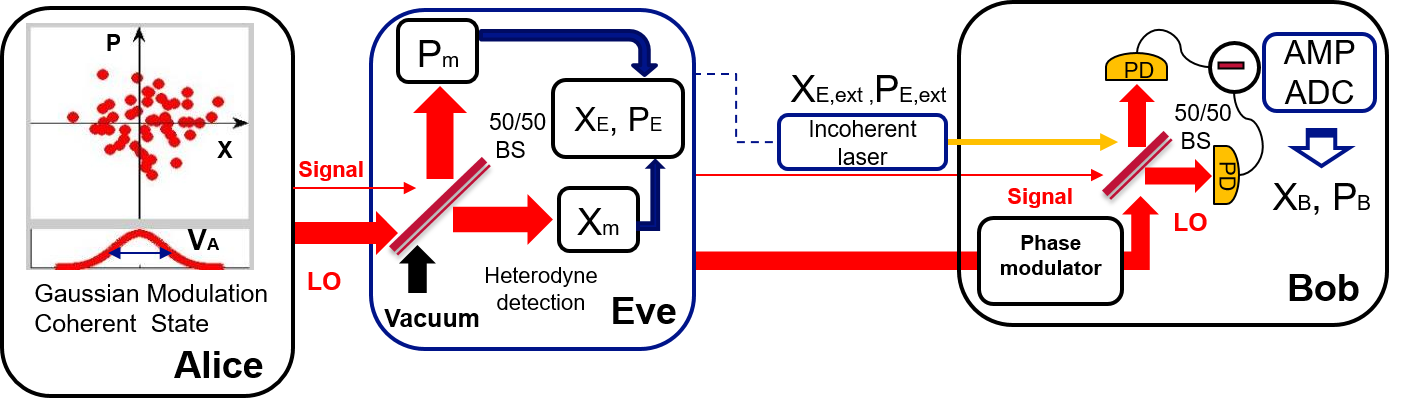}}
	\caption{ A concept scheme of Eve's HD blinding attack in CV-QKD. Alice: preparation of the coherent state with Gaussian modulation ($V_A$); Eve: Heterodyne measurement $X_M, P_M$, re-preparation $X_E, P_E$ and external light state $X_{E,ext}, P_{E,ext}$; Bob: HD measurement:$X_B, P_B$; BS: beam-splitter, LO: local oscillator, PD, pin photodiode, AMP: amplifiers, ADC: analog-to-digital converter}
	\label{eve}
\end{figure*}

\label{attack}
\subsection{Eve's attack strategy} \label{EAS}
By targeting on a typical implementation of CV-QKD \cite{Jouguet2013a}, we now present Eve's HD blinding attack strategy step by step along with a realistic implementation of Alice and Bob's GMCS CV-QKD protocol (a concept scheme of the attack strategy is shown in \cref{eve}):

\begin{enumerate}
	\item In GMCS CV-QKD, Alice prepares quantum signal $|X+iP\rangle$ in which amplitude $X$ and phase $P$ quadratures are continuously modulated with a bivariate centered Gaussian distribution (Due to the symmetric of $X$ and $P$, we will only look at the $X$ quadrature in our analysis):
 \begin{align}
	\label{xb2}
	X=X_A+X_0,
	\end{align}
	 with a variance $V_{A}=\langle X^2_{A} \rangle-\langle X_{A} \rangle^2$ and the vacuum noise $\langle X^2_{0} \rangle=1$. 
	
\item Eve cuts down the quantum channel and performs a full IR attack \cite{Lodewyck2007a} : Eve intercepts Alice's signals by performing heterodyne detection on $X$ and $P$ quadratures to obtain measurement result: 
\begin{align}
X_M=\dfrac{1}{\sqrt 2}(X_A+X_0+X_0'),
\end{align}
here, due to the BS in heterodyne detection, there is a factor $1/\sqrt 2$ for loss and a vacuum state $X_0'$ added. According to her measurements ($X_M, P_M$), Eve prepares and resends her noisy coherent states $|X_E+iP_E\rangle$ as signals to Bob through a lossy channel with transmission of $T$:
\begin{equation}
\label{ve}
\begin{split}
X_E =&gX_M+X_0''\\
=&\dfrac{g}{\sqrt 2}(X_A+X_0+X_0')+X_0'',
\end{split}
\end{equation}
in which, $g=\sqrt 2$ is the gain factor to compensate the loss due to heterodyne detection and $X_0''$ is a noise term due to coherent state encoding of Eve. $X_0'$ and $X_0''$ all follow a centered normal distribution with unity variance.

\item Along with her resending signals, Eve inserts external laser pulses into the signal port of Bob's HD which is not coherent with CV-QKD signals. In practice, Eve needs to choose the properties of the external light: wavelength is slightly different from Alice's signal; pulse width and repetition rate are same as Alice's signals; intensity or photons per pulse depends on how much Eve wants to influence on Bob's HD measurement which will be analyzed in the next section. In order to insert such pulses into Bob's HD, Eve can set the polarization of them as the ones of the CV-QKD signals, if polarization multiplexing technique is used \cite{Jouguet2013}.
	
\item Bob performs   HD measurements on the incoming Eve's signal pulses interfering with LO pulses and the external laser pulses interfering with vacuum. Since the external laser pulses are incoherent with CV-QKD signals, there are no interferences between the external light and LO pulses, we can independently analyze the impact of external laser and Eve's IR signals on Bob's HD output signals. Regarding to a HD with an ideal infinite detection range ($-\infty, \infty$) and an efficiency of $\eta$, Bob's HD output signal can be given as:
\begin{equation}
\label{xbi}
\begin{split}
X_{Bi}&=\sqrt{\eta I_\text{lo}}[\sqrt{\eta T}(X_E+X_\text{tech})+\sqrt{1-\eta T}X_0''']\\ &+X_{E,ext}+X_\text{ele},
\end{split}
\end{equation}

in which  $X_\text{tech}$ is the noise term due to any technical noises from Eve, Alice and Bob's devices; $X_0'''$ is another vacuum state due to loss in Bob's HD; $X_{E,ext}$ is the external light state that impacts on Bob's HD output, which 
can be treated as the case there is only LO pulses go into HD (\cref{xhd} in \cref{HDtheo}):
\begin{align}
\label{xbhd}
X_{E,ext}=\eta (1-2T_\text{ext})I_\text{ext}+2\sqrt{\eta T_\text{ext}(1-T_\text{ext})I_\text{ext}}X_0'''',
\end{align}

here $T_\text{ext}$ is the overall transmission of HD regarding to the external light pulses that goes into signal port, $I_\text{ext}$ is number of photons per external light pulse, $X_0''''$ is the vacuum state that interferes with the external light. As Bob's HD balance setting is only valid for LO pulses go into LO port, the overall imbalance factor $\epsilon_\text{ext}=1-2T_\text{ext}$ of the external light pulses will contribute non-negligible offsets to final HD signals. The second term of \cref{xbhd} will contribute its own shot noise into CV-QKD excess noise. In a realistic case, Bob's HD only operates linearly with a finite detection range [$\alpha_2, \alpha_1$] as discussed in \cref{seciii}, thus Bob's HD output signal is given by \cref{xbs} in which $X_{Bi}$ [\cref{xbi}] replaces the term of $X_{HD}$.

\item Alice and Bob perform classical post processing on their correlated data ($X_A, X_B$): sifting, parameter estimation, reverse reconciliation and privacy amplification in order to obtain keys. Due to Eve's external light, Alice and Bob may believe their excess noise is still below the null threshold which will cause them to accept compromised keys.

% In order to achieve such security breach , Eve needs to properly set $I_\text{ext}$ of the external laser to effectively bias the noise due to the IR attack and the external light, which will be analyzed in the next section.

\item Eve listens to the classical communication between Alice and Bob, in order to perform the same post processing of Alice and Bob on her data to get identical keys. 
\end{enumerate}

\section{Security analysis and simulations} \label{secv}
In this section, we will demonstrate in simulation of Eve's attack strategy in \cref{EAS} and show that how Eve can in practice break the security of Alice and Bob GMCS CV-QKD system with a realistic parameter setting.

% We will show that our attack is feasible towards a typical implementation of GMCS CV-QKD \cite{Jouguet2013a} and Eve's power is restricted to the current technology level. 

\subsection{Realistic assumptions of Alice, Bob and Eve}

In the security analysis, it is necessary to assume Alice and Bob's CV-QKD implementation setup; and Eve's power in a realistic scenario such that Eve's security breach can be valid. We first consider the assumptions of Alice and Bob's CV-QKD implementation and their device parameters:

\begin{itemize}
	\item Alice optimizes her Gaussian modulation variance $V_A \in \{1,100\}$ based on the distance \cite{Jouguet2011}.

	\item Bob balances his HD on the LO pulses that go into LO port such that $T_\text{lo}\approx0.5$ and one LO pulse contains $I_\text{lo}=10^8$ number of photons at Bob side. Thus the impact of LO leakage on HD is assumed to be negligible. 
	
	\item Alice and Bob implement real time shot noise calibration as in Ref. \cite{Jouguet2013}, their shot noise $N_0=\eta I_\text{lo}$ is assumed to be not tampered by Eve. Such assumption can be extended to the case that Alice and Bob use LLO scheme \cite{Qi2015}
		
	\item Bob's HD efficiency $\eta=0.6$, electronic noise variance $v_\mathrm{ele}=0.01 N_0$, linear detection limit $\alpha_1=-\alpha_2=20 \sqrt{N_0}$. Such limits are considered large enough to ensure Bob's HD operating in normal case. Alice and Bob calibrate $\eta$ and $v_\mathrm{ele}$ before CV-QKD protocol.

 \item Alice and Bob perform reverse reconciliation with a reconciliation efficiency of $95\%$.
	
\end{itemize}
	
We now consider Eve's attack strategy assumptions:
\begin{itemize}

\item Eve's station is right after Alice station. The loss between Alice and Bob is identical to the one between Eve and Bob which is given by $T=10^{-aL/10}$, $L$ is the distance between Alice and Bob, $a=0.21 \deci \bel/\kilo \meter$ is the standard loss coefficient of single mode fiber in 1550 nm.
		
\item Eve inserts the external light beside Bob's station such that Eve can control precisely its power ($I_\text{ext}$) without it going through the lossy channel. 
	
\item Eve inserts the external light into Bob's HD signal port and its overall transmission on Bob's HD is $T_\text{ext}=0.49$. Note Eve is assumed to know $T_\text{ext}$ and can control its value by using shorter or longer wavelength as in the wavelength attack \cite{Ma2013a,Huang2013,Huang2014}. 
	
\end{itemize}

\subsection{Eve's impact and excess noise contribution}
Based on these assumptions, we can now analyze Eve's impact and excess noise contribution over Alice and Bob CV-QKD protocol. From Alice and Bob's point of views, all the statistical quantities need to be normalized into shot noise units which will be considered in the following analysis. From Eve's strategy mentioned above, there are mainly three parts of excess noise due to Eve's attack: noise due to IR attack $\xi_{IR}$, noise due to the external light $\xi_\text{ext}$ and noise due to technical imperfections $\xi_\text{tech}$. Since LO pulses and external light pulses are in different modes, we can separately evaluate $\xi_{IR}$ and $\xi_\text{ext}$.

As in the step (2) of Eve's strategy, Eve's IR attack adds one vacuum noise due to the 50/50 BS in the heterodyne detection ($X_0'$) and another one due to coherent encoding ($X_0''$) which gives $\xi_{IR}=2$ \cite{Lodewyck2007a}. We further consider total technical noise due to Alice, Bob and Eve devices imperfections as $\xi_\text{tech}=0.1$ which is an experimental result in Ref.~\cite{Lodewyck2007a}. 

\begin{figure}
	{\includegraphics[width=0.5\textwidth]{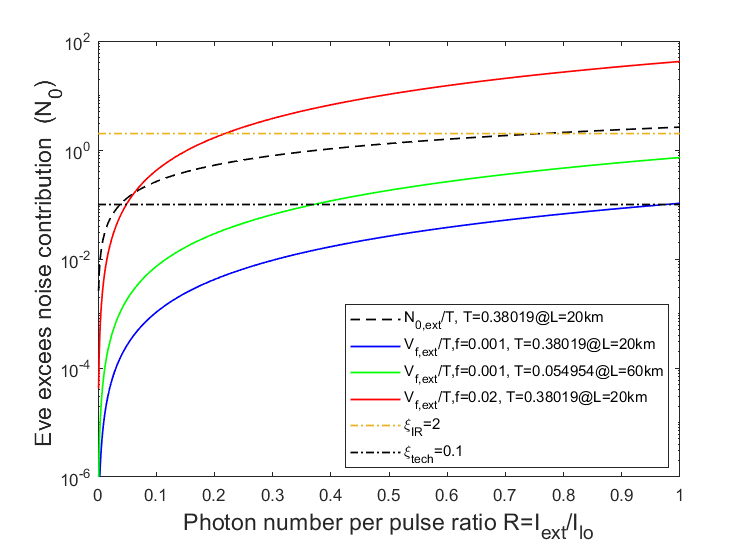}}
	\caption{Excess noise contributions, for different blinding attack parameters, as a function of the photon number per pulse ratio $R$. Solid curves stand for the excess noise due to blinding laser intensity fluctuation (see text). Dashed curve stands for excess noise added by blinding laser shot noise. The upper dashed dotted line stands for the excess noise due to the IR attack, the lower dashed dotted line stands for the technical excess noises, which are independent of the external blinding laser.}
	\label{noise}
\end{figure}

Regarding to noise due to the external light in the step (3) and (4) of Eve's strategy, as the analysis in \cref{HDtheo}, there are mainly two parts: the external light's own shot noise $N_\text{0,ext}$ and laser intensity fluctuation noise $V_\text{f,ext}$ due to insufficient subtraction of Bob's HD. If we further express these values into shot noise units with $\eta I_\text{lo}$, we can know their excess noise contribution: 
\begin{align}
\label{n0e}
N_\text{0,ext}= 4T_\text{ext}(1-T_\text{ext})I_\text{ext}/I_\text{lo}, \\
\label{vfe}
V_\text{f,ext}= \eta f_\text{ext}^2(1-2T_\text{ext})^2I_\text{ext}^2/I_\text{lo},
\end{align}
in which $f_\text{ext}$ is external laser's intensity fluctuation ratio, we  consider that Eve has an ultra stable laser source $f_\text{ext}=0.1\%$ or a normal laser source $f_\text{ext}=2\%$. Thus the total noise due to Eve's external light is given by:
\begin{align}
\label{vb2}
V_{B2}&=N_{0,ext}+V_{f,ext} \\
&= 4T_\text{ext}(1-T_\text{ext})R+ R^2 \eta f_\text{ext}^2(1-2T_\text{ext})^2I_\text{lo},
\end{align}
in which $R=I_\text{ext}/I_\text{lo}$ is the ratio between photon number of one Eve's external light pulse and one Bob's LO pulse. Note $V_{B2}$ is the noise due to external light at Bob side, the equivalent noise of $V_{B2}$ on Alice side needs take the transmission $T$ into account. Thus the total excess noise due to the external light is given as:
\begin{align}
\xi_\text{ext}=4T_\text{ext}(1-T_\text{ext})R/T+ R^2 \eta f_\text{ext}^2(1-2T_\text{ext})^2I_\text{lo}/T.
\end{align}
We now summarize all these noises due to Eve's attack in \cref{noise}. As we can see, $N_\text{0,ext}$ and $V_{f,ext}$ due to external laser increases with $I_\text{ext}$. If Eve uses a stable laser source as her external light with $f_\text{ext}=0.1\%$, the dominant noise contribution is from its shot noise $N_\text{0,ext}$. However if Eve uses a common laser source with $f_\text{ext}=2\%$, the intensity fluctuation noise $V_\text{f,ext}$ will take the lead and induce more disturbances on CV-QKD signals, which  needs to consider in practice. In our later analysis, we will consider $f_\text{ext}=0.1\%$ in Eve's attack. We can also observe that Eve's external light noise increases with Alice and Bob distance $L$ due to the factor of $1/T$, as the external light is inserted at Bob's side. 
On the other hand, Eve's external light also contributes a non-negligible offset on Bob's HD output signal as discussed in \cref{HDtheo}, which is under Eve's control through $T_\text{ext}$ and $I_\text{ext}$:
\begin{align}
\label{d}
D_\text{ext}=\sqrt{\eta/I_\text{lo}}(1-2T_\text{ext}) I_\text{ext} =R\sqrt{\eta I_\text{lo}}(1-2T_\text{ext}).
\end{align} 
Note $D_\text{ext}$ is normalized in $\sqrt{N_0}$. As $D_\text{ext}$ is proportional to $I_\text{ext}$, it means if Eve wants more influence from external light on Bob's HD, she needs to increase $\xi_\text{ext}$ which may potentially limit the power of the attack. In order to achieve a security breach, Eve needs to properly set $D_\text{ext}$ in order to cause large enough offset to force Bob's HD works in the saturation region, which will help Eve to effectively bias the noises $\xi_{IR}$, $\xi_\text{ext}$ and $\xi_\text{tech}$ due to the attack. 
 
\subsection{Alice and Bob's parameter estimation under Eve's attack}
In order to determine whether Eve can have a security breach under the HD blinding attack, we need to evaluate the parameter estimation of Alice and Bob: channel transmission $\hat T$ and excess noise $\hat \xi$, to see whether  Eve can bias the excess noise due to the attack small enough such that Alice and Bob believe they can still share a secret key. A security breach thus corresponds to the condition: $\hat \xi<\xi_\text{null}$, in which $\xi_\text{null}$ is the null key threshold corresponds to the maximum excess noise that allows Alice and Bob to extract a secret key under collective attack model\cite{Navascues2006} for given values of $\hat T$ and $V_A$. According to the standard parameter estimation procedure of CV-QKD in \cref{GMCS}, we can estimate $\hat T$ and $\hat \xi$ based on \cref{eq2,eq3}.

We first consider the case where Bob's HD linear range is infinite ($-\infty, \infty$). In this case, we can predict the mean of Bob's HD measurement $\langle X_{Bi} \rangle=D_\text{ext}$ and its variance:
\begin{align}
\label{vbi}
V_{Bi} = &\langle X^2_{Bi} \rangle-\langle X_{Bi} \rangle^2 \\
=& \eta T (\xi_{IR}+\xi_\text{ext}+\xi_\text{tech})+ 1 + v_\mathrm{ele},
\end{align}
 in which, we can directly deduce the channel transmission estimation of Alice and Bob: $\hat T_\text{i}=T$ and their excess noise estimation: $\hat \xi_\text{i}=\xi_{IR}+\xi_\text{tech}+\xi_\text{ext}$. It is obvious that Alice and Bob can easily spot Eve's attack action if Bob's HD works in linear region, as $\hat \xi_\text{i}>>\xi_\text{null}$ at any distances. 
\begin{figure}
	{\includegraphics[width=0.5\textwidth]{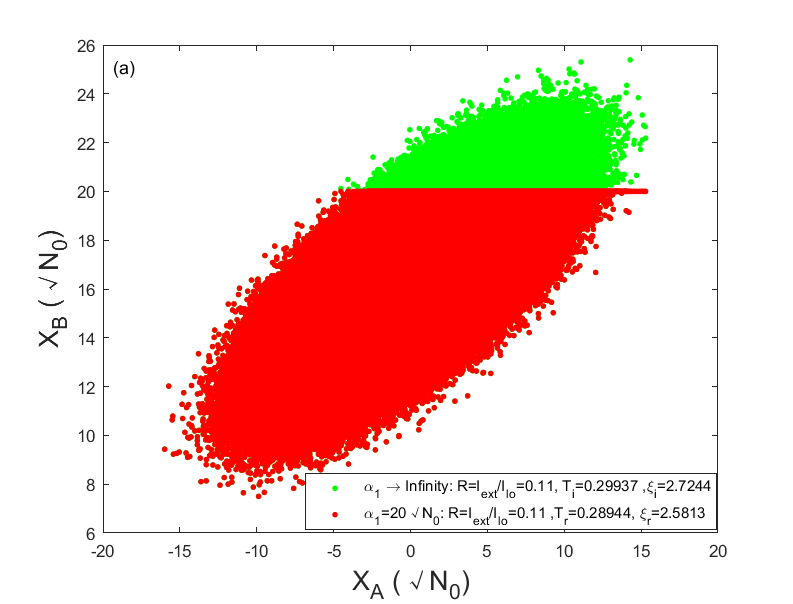}}
	% \hfill
	{\includegraphics[width=0.5\textwidth]{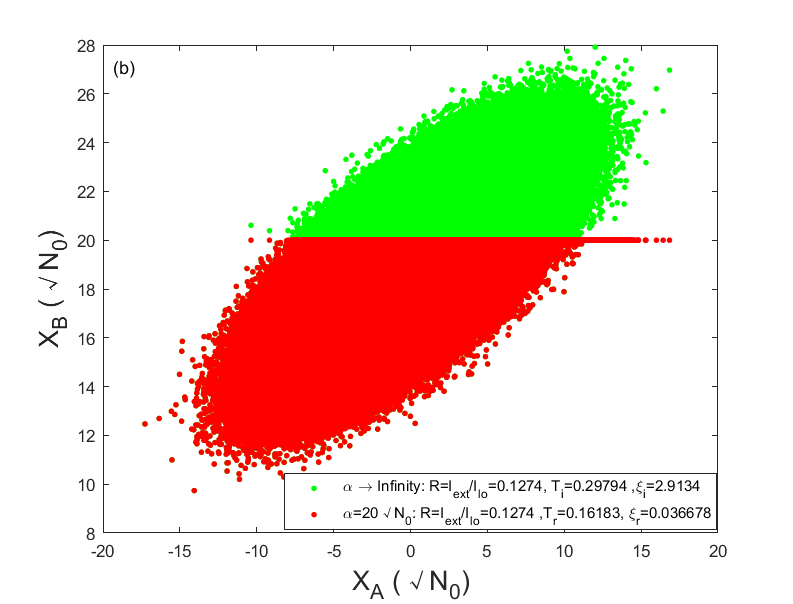}}
	\caption{Simulation of the impact of  detector saturation  on the quadratures distribution  $X_B$ versus $X_A$, for two sets of parameters. Red (dark gray): $X_{B}$ HD detection range $[-20\sqrt{N_0},20\sqrt{N_0}]$, green (light gray): $X_{B_{i}}$ HD ideal linear case. The corresponding estimation of $\hat T$ and $\hat \xi$ are both  given in the legend for $X_{B}$ and $X_{Bi}$ respectively. (a) $R=0.11$,  no security breach as $\hat \xi_\text{r} >>\xi_\text{null}$; (b) $R=0.1274$, security breach as $\hat \xi_\text{r} <\xi_\text{null}$. For both case, $\hat \xi_\text{i}>>\xi_\text{null}$.}
	\label{xb}
\end{figure}
However, if Bob performs a realistic HD measurement with a finite linear range [$\alpha_2, \alpha_1$], Eve can manipulate Bob's HD signal statistics by controlling  $D_\text{ext}$ through $R$ which affect Bob's HD output statistics and further bias Alice and Bob's parameter estimation. We now demonstrate Eve's action on $R$ and its impact on Alice and Bob's data ($X_A, X_B$) in simulation. As shown in \cref{xb} (a) and (b) , we consider Eve uses $R=0.1$ and $R=0.1274$ respectively in her strategy at distance $L=25 \kilo \meter$ , in which $X_A, X_{Bi}$ (green or light gray) correspond to Bob's HD linger range is infinite ($-\infty, \infty$) and ($X_A, X_{B}$) (red or dark gray) correspond to Bob's HD is limited to [$\alpha_2, \alpha_1$]. If Alice and Bob perform parameter estimation based on ($X_A, X_{Bi}$), there is only a displacement $D_\text{ext}$ has been introduced on Bob's data, they can still notice Eve's attack base on $\hat T_\text{i}$ and $\hat \xi_\text{i}$. However, Alice and Bob may not be able to detect Eve's action based on ($X_A, X_{B}$), as Eve can gradually increase $R$ in order to force parts of Bob's HD signals saturated and bias the estimation $\hat T_\text{r}$ and $\hat \xi_\text{r}$. Due to Bob's HD saturation,  $X_{B}$ variation is limited by Bob's upper detection limit $\alpha_2=20$ which results in smaller variance of Bob compared to the one of $X_{Bi}$ and a weaker covariance correlation between Alice and Bob, which leads $\hat T_\text{r}< \hat T_\text{i}$. If Eve chooses properly the value of $R$, she can eventually meet the condition $\hat \xi_\text{r}<\xi_\text{null}$. In \cref{xb}(a) Eve's choice of $R=0.1$ can not lead to a security breach as $\hat \xi_\text{r}=2.5813 >\xi_\text{null}=0.1013$ at $25 \kilo \meter$. If Eve keeps increasing $I_\text{ext}$, as shown in \cref{xb}(b) the choice of $R=0.1274$ corresponds to a security breach condition as $\hat \xi_\text{r}= 0.0367 <\xi_\text{null}$. In \cref{xb}, statistical measurements are based on $N=10^7$ simulation data for linear HD (green or light gray) and saturation HD case (red or dark gray). It shows that Eve's external light  power needs to be high enough to affect sufficiently Bob's data distribution in order to achieve a security breach, otherwise, Alice and Bob can still detect the noise due to Eve's attack.

\begin{figure}
	{\includegraphics[width=0.5\textwidth]{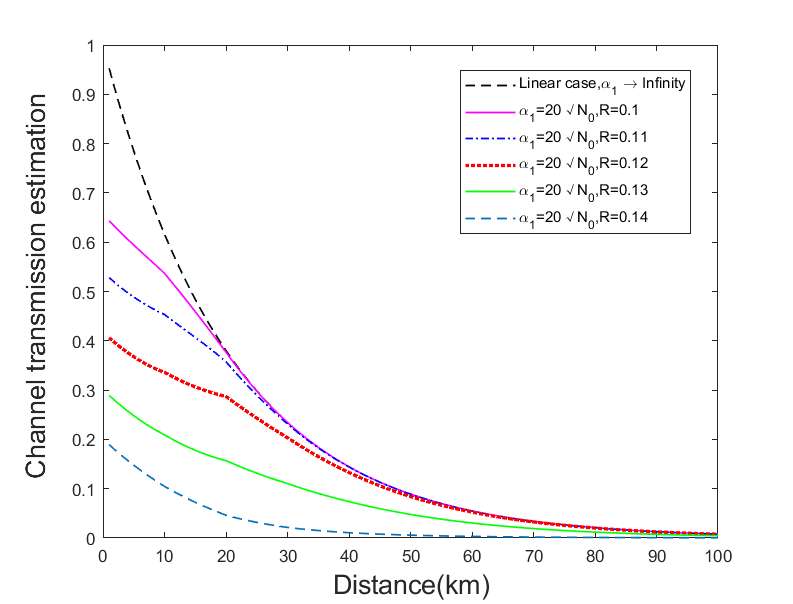}}
	\caption{Alice and Bob's transmission estimation versus distance under Eve's HD blinding attack. The black dashed curve (top one) corresponds to  $\hat T_\text{i}$ that is estimated by $X_{Bi}$ in the linear case ; while the other lower curves correspond to  $\hat T_\text{r}$ in the realistic case  that are estimated by $X_B$ with $R=0.1, 0.11, 0.12, 0.13, 0.14$ from top to bottom.}
	\label{T}
\end{figure}

\begin{figure}
	{\includegraphics[width=0.5\textwidth]{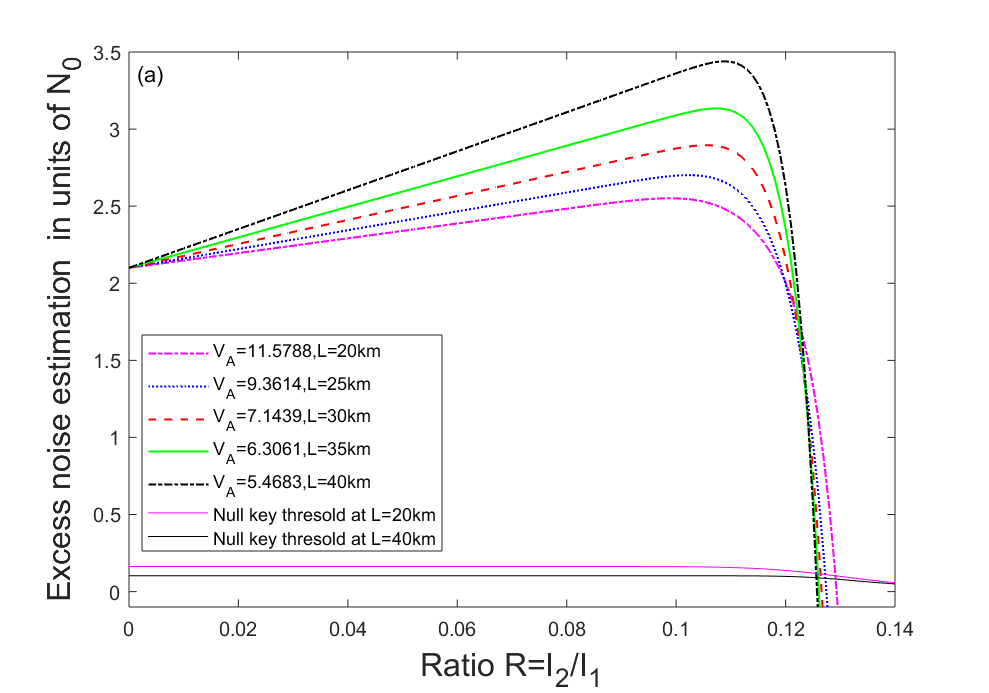}}
	% \hfill
	{\includegraphics[width=0.5\textwidth]{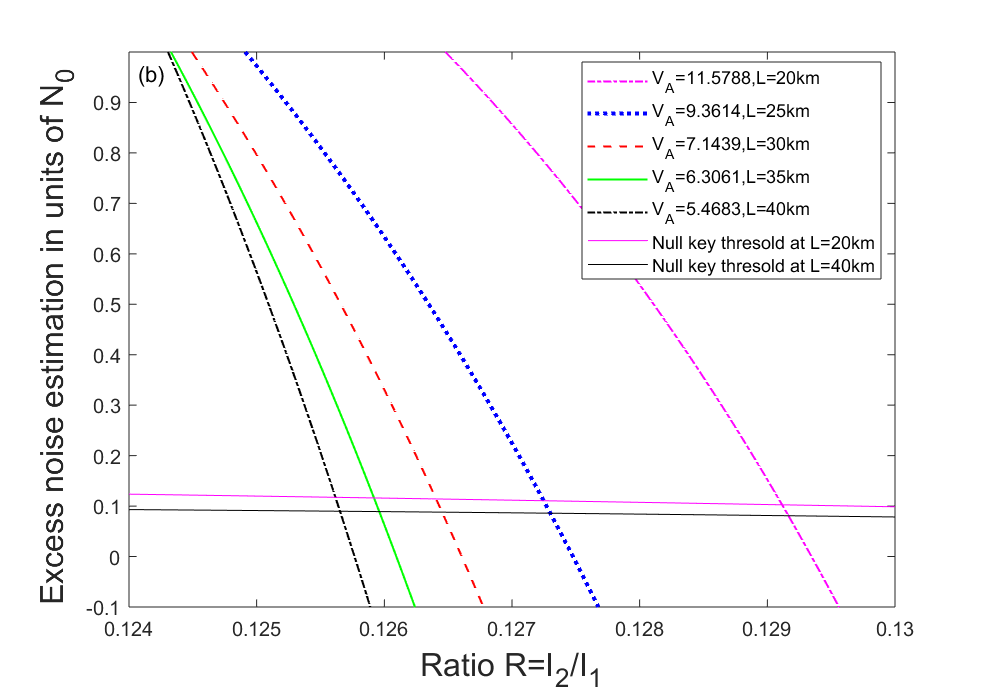}}
	\caption{Excess noise estimation of Alice and Bob versus photon number ratio $R$, $\eta=0.55$, $T_\text{ext}=0.49$, $f_\text{ext}=0.1\%$, $v_\mathrm{ele}=0.01$. (a) Rang of $R$: 0-0.14,  (b) Range of $R$: 0.124-0.13. The upper five curves stand for the excess noise estimations $\hat \xi_\text{r}$ at different distances, the lower two curves stand for the null key thresholds $\xi_\text{null}$ at 20 \kilo \meter and 40 \kilo \meter.}
	\label{n}
\end{figure}

We further analyze Eve's choice of $R$ to meet the condition $\hat \xi_\text{r} <\xi_\text{null}$. In the simulation of Eve's attack, we use the HD model in \cref{HDtheo} and standard parameter estimation procedure of CV-QKD in \cref{GMCS} to estimate $\hat T_\text{r}$ and $\hat \xi_\text{r}$ for Alice and Bob. Particularly, we calculate $\hat T_\text{r}$ and $\hat \xi_\text{r}$ by increasing the value of $I_\text{ext}$ and thus the ratio $R$. In \cref{T}, we show the impact of $R$ on $\hat T_\text{r}$ over distance $L=0\sim100 \kilo \meter$. 
As shown in \cref{T}, Eve's external light reduce Alice and Bob's channel transmission $\hat T_\text{r}$ as expected, however such reduction will not prevent Alice and Bob to proceed to key generation. As long as  $\hat \xi_\text{r}<\xi_\text{null}$, there is still a security breach.  To illustrate Eve's impact on Alice and Bob's estimation of $\hat \xi_\text{r}$, we continuously increase $R$ and  deduce corresponding $\hat \xi_\text{r}$ and $\xi_\text{null}$ for a given setting of $V_A$ and $\hat T_\text{r}$. The  results of  $\hat \xi_\text{r}$ and $\xi_\text{null}$ versus $R$ are shown in \cref{n} for different distances $L=20, 25, 30, 35, 40 \kilo \meter$, in which excess noise $\hat \xi_\text{r}$ in HD linear region increases with $R$  and with distance due to the factor of $\hat T_\text{r}$. According to the previous analysis, Eve's noises consist constant noises: $0.1$ due to technical imperfections and $2$ due to IR attack; variable noises increasing with $R$: the shot noise of the external laser and its intensity fluctuation noise. The total noise is much higher than the tolerable excess noise for Alice and Bob to generate a key ($\xi_\text{null}\sim 10\%N_0 $) and thus it will reveal Eve's presence. 

However $\hat \xi_\text{r}$ decrease sharply when $R>0.12$, since corresponding offset $D_\text{ext}$ overpass the HD detection limit $\alpha_1$, such that $\hat \xi_\text{r}$ is effectively biased by Eve. As shown in previous analysis, due to HD saturation, Bob's HD variance and his data's covariance with Alice both become smaller. However the impact of HD saturation on its variance degradation is much larger than on the covariance, which results in a quick drop of $\hat \xi_\text{r}$. Although the curves in \cref{n}(a) sharply decreases around $R=0.12$, each value of $\hat \xi_\text{r}$ only corresponds to one value of $R$. As shown in \cref{n}(b), a more precise control of $R$ will help Eve to manipulate $\hat \xi_\text{r}$ to an arbitrary value between $0$ and $\xi_\text{null}$. It means once Eve has enough precision on the power of the external laser, she can accurately manipulate Alice and Bob's excess noise estimation to any small value she desires. For example, according to the simulations, a successful attack is possible with the choice of $I_\text{ext}=RI_\text{lo}=0.1274\times10^8=1274\times10^4$ and $f_\text{ext}I_\text{ext}=1274$ which shows that Eve needs a precision of $10^4$ photons and a stability of $10^3$ photons level on one external laser pulse in order to accurately bias the excess noise estimation. Such precision is realistic and achievable with current technology. 

For a given distance, Eve can in practice choose a proper value of $R$ to achieve $\hat \xi_\text{r} <\xi_\text{null}$ such that Alice and Bob still believe they share a secure key according to their parameter estimation and proceed to key generation however the generated keys are not secure at all because of Eve's IR attack. In principle, Eve can set $\hat \xi_\text{r}$ to be arbitrary close to zero, which further enables her to control Alice and Bob key rate generation. \Cref{n} is a reference for Eve to properly set the value of $R$. 

\section{Countermeasures}\label{secvi}
In DV-QKD, blinding attack is well known for breaking security on various protocols through controlling different types of SPDs such as avalanche photodiode (APD) \cite{Makarov2009,Lydersen2010} and superconducting nanowire SPD \cite{Lydersen2011,Lydersen2011a}. This kind of detector controlled attack based on bright illumination now extends to CV-QKD using HD as shown in this paper. In both CV and DV case, Eve is required to send a relatively strong classical light to actively control Bob's detector. Due to such similarity, countermeasures against blinding attack in DV-QKD are thus worth to consider to defeat HD blinding attack in CV-QKD. Here we briefly discuses several possible countermeasures against HD blinding attack in CV-QKD and compare them with the ones in DV-QKD.
 
In the first approach, a straight forward countermeasure is to monitor the light intensity that is going into signal port of HD. Bob can implement such countermeasure using a sensitive p-i-n photodiode, in order to detect any strong light impinging on the signal port. Such method can be also used for energy test that is required in some security proofs \cite{Leverrier2017}. It is however challenging to build a detection system that can give in practice the capability to detect light in any optical mode that Eve ay try to use. In DV-QKD, such watchdog detectors have been proposed \cite{Gerhardt2011} and implemented \cite{Dixon2017} to detect blinding attack. However, it has been shown that practical watchdog may not always be able to rise the alarm \cite{Sajeed2015}. Moreover, Eve may also use high power laser to damage the photodiode watchdog and bypass the security alarm of the QKD protocol \cite{Makarov2016}.

Bob's HD  consists of two classical photodiodes (PDs). Hence, instead of measuring the difference between the two photocurrents, Bob may also monitor one, or the sum of the two photocurrents, from two PDs. This may be however challenging in practice, as it will disturb the HD output, and as it can be difficult to set a proper discrimination level to correctly detect, against blinding attack. In our proposed blinding attack strategy, Eve only need to increase the overall energy by about 12\%. Hence LO intensity fluctuations, due to Alice's laser source and channel environment may exceed the external light's energy, which can lead to false alarms. In addition, it is currently been technically challenging to design a high gain, high bandwidth, high efficiency, low noise HD in practice for CV-QKD purposes. Adding extra electronics components can increase the electronic noises and reduce HD performance. A similar approach in DV-QKD has been also proposed, in which Bob monitors the photocurrent from APD \cite{Yuan2010}. Unfortunately, such method was later proven not sufficiently to detect the blinding attack in many particular cases \cite{Lydersen2010a}. Moreover, in addition to  monitoring the photocurrent from APD, one may use the synchronization detector as an auxiliary monitor to detect the blinding light \cite{Wang2014}.
 
A third countermeasure has been proposed in the Ref.\cite{Kunz-Jacques2015}: Alice and Bob test the linearity between the noise and signal measurement by using an active attenuation device on Bob's side, i.e. an amplitude modulator. Such method explores the linearity of HD and  can thus in principle prevent the HD blinding attack: the randomization of signal port's attenuation prevents Eve to properly set the intensity of blinding pulses. However, a practical amplitude modulator is wavelength sensitive which can loose its amplitude extension when the wavelength is out of the spectral range. In addition, such linearity test increases the implementation complexity and detector losses. A similar approach, based on random detector efficiency  has also been proposed in DV-QKD \cite{Lim2015}, yet it has been shown that it is not always effective in a practical implementation \cite{Huang2016a}.
 
The three previous approaches require some modifications to the CV-QKD system hardware, leading to additional experimental complexity. We suggest, on the other hand, that data post-processing combined with calibrated homodyne detection, can be a simple and efficient way to counter the blinding attack. We propose the following generic method: Bob sets security thresholds [$S_2, S_1$] inside the HD limits  [$\alpha_2, \alpha_1$]. In the parameter estimation stage, Alice and Bob can thus estimate, for each block, the fraction of the HD measurement data that have been recorded outside of the interval [$S_2, S_1$]. If a too large fraction of HD measurements is recorded beyond the thresholds, then Alice and Bob know the HD was not working in its linear range for some non neglectable fraction of the quantum communication phase, and they discard the block. This approach relies on the per-calibration of HD detection limits (the values of $\alpha_{1}$ and $\alpha_{2}$ which is required to be performed with a good precision (compared to $N_0$) and in a safe environment. Setting the value of the confidence interval  [$S_2, S_1$] and the fraction of rejected data that can be tolerated will require further work taking finite-size effects into account and including a characterization of statistical fluctuations.
 
Finally, since HD blinding attack is a detector-based attack, MDI CV-QKD \cite{Li2014,Ma2014a,Pirandola2015} can be a potential solution to defeat such attack. Although a proof-of-principle demonstration of MDI CV-QKD has been already performed in experiment \cite{Pirandola2015}, there is still a large gap between practical implementation and theoretical proposal. There are even debates on whether MDI CV-QKD can become practical regarding to its theoretical performance limitations and current available technologies \cite{Xu2015,Pirandola2015a}. Recent works in finite size \cite{Papanastasiou2017} and composable \cite{Lupo2018} security proofs of MDI CV-QKD have shown some practical feasibilities of such protocol from  theoretical perspectives. This paper may be an additional motivation for future development of practical MDI  in CV-QKD, similarly to the role played by the blinding attack in DV-QKD, to trigger the birth of MDI QKD \cite{Braunstein2012, Lo2012} and its deployment \cite{Yin2016}. 
 
These countermeasures show that current CV QKD implementations need some upgrades in hardware or software to defeat the proposed attack. It is even more important to verify the functionality of these countermeasures in practice, as they may fail to defeat the attacks if they are not correctly implemented as in the cases of DV-QKD \cite{Huang2016a,Lydersen2010a}.

\section{Conclusion} \label{secvii}
\label{conclusion}

In this paper, we detail an attack strategy exploiting the homodyne detection vulnerability \cite{Qin2016,Qin2013} that is moreover implementable with low experimental complexity. Inspired by and analogous to the blinding attack in DV-QKD, our attack allows Eve to influence Bob's homodyne detection response, by sending external light. We demonstrate that this attack can constitutes a powerful strategy, that can fully break the security of practical CV-QKD systems. Based on experimental observations, we propose an effective model to account for homodyne detection imperfections and use it to model Eve's attack. Simulation results illustrate the feasibility of our proposed attack under realistic experimental conditions. Compared to other side channel attacks in CV-QKD requiring complex experimental techniques  \cite{Qin2013, Huang2014,Huang2013}, we believe our strategy should be simple enough to allow effective eavesdropping demonstration on deployed CV-QKD. This attack hence highlights the importance of exploring the assumptions in security proofs when implementing CV-QKD protocols and the necessity to implement suitable countermeasure to ensure the practical security of CV-QKD systems.

\acknowledgments
This work was funded by NSERC of Canada (programs Discovery and CryptoWorks21), CFI, MRIS of Ontario, French National Research Agency (ANR Emergence project Quantum-WDM), European Commission (Marie Sk\l{}odowska-Curie ITN project QCALL), and Quantum Communications Hub through EPSRC UK National Quantum Technology Programme.

\end{document}